\documentclass[physics,article,accept,oneauthor,pdftex]{Definitions/mdpi} 

\firstpage{1} 
\pubvolume{1}
\issuenum{1}
\articlenumber{0}
\pubyear{2021}
\copyrightyear{2021}
\datereceived{16 September 2021 } 
\dateaccepted{05 November 2021 } 
\datepublished{} 
\hreflink{https://doi.org/} 

\Title{A Shock-in-Jet Synchrotron Mirror Model for Blazars}

\TitleCitation{A Shock-in-Jet Synchrotron Mirror Model for Blazars}

\Author{{Markus B\"ottcher} \orcidA{0000-0002-8434-5692}}

\AuthorNames{Markus B\"ottcher}

\AuthorCitation{B\"ottcher, M.}

\address[1]{%
Centre for Space Research, North-West University, Potchefstroom 2520, South Africa; Markus.Bottcher@nwu.ac.za
}

\abstract{Reinhard Schlickeiser has made groundbreaking contributions to various aspects of blazar physics, including diffusive shock
acceleration, the theory of synchrotron radiation, the production of gamma-rays through Compton scattering in various astrophysical
sources, etc. This paper, describing the development of a self-consistent shock-in-jet model for blazars with a synchrotron mirror 
feature, is therefore an appropriate contribution to a Special Issue 
in honor of Reinhard Schlickeiser's 70th birthday. The model
is based on our previous development of a self-consistent shock-in-jet model with relativistic thermal 
and 
non-thermal particle distributions
evaluated via Monte-Carlo simulations of diffusive shock acceleration, and time-dependent radiative transport. This model has been 
very successful 
in modeling spectral variability patterns of several blazars, but has difficulties describing orphan flares, i.e., high-energy flares
without a significant counterpart in the low-frequency (synchrotron) radiation component. As a solution, this paper investigates the
possibility of a synchrotron mirror component within the shock-in-jet model. It is demonstrated that orphan flares result naturally in
this scenario. The model's applicability to a recently observed orphan gamma-ray flare in the blazar 
3C279 is discussed and it is found that only
orphan flares with mild ($\lesssim$ a factor of 2--3) enhancements of the Compton dominance can be reproduced in a synchrotron-mirror
scenario, if no additional parameter changes are invoked. }

\keyword{active galaxies; blazars; diffusive shock acceleration; radiative transport; gamma-rays} 

\begin{document}

\section{Introduction}

Blazars are a class of jet-dominated active galactic nuclei. As~most convincingly argued by Reinhard Schlickeiser (RS)
in 1996~\cite{Schlickeiser96}, their broad-band non-thermal emission, ranging from radio to gamma-rays, must
be strongly Doppler boosted due to relativistic motion of an emission region along the jet, oriented close to our
line of sight. The~spectral energy distributions (SEDs) of blazars are dominated by two broad, non-thermal radiation
components. The~low-frequency component, from~radio to optical/UV/X-ray frequencies, is generally attributed to
synchrotron radiation by relativistic electrons. Most notably, Crusius and Schlickeiser~\cite{CS86,CS88} have evaluated
the angle-averaged synchrotron emission from isotropically distributed electrons in random magnetic fields, including
plasma effects, which are now frequently used as the standard expressions for the low-frequency emission from
blazars. However, note also an alternative suggestion by RS in 2003~\citep{Schlickeiser03} that the low-frequency 
emission may be produced as electrostatic bremsstrahlung, i.e.,~the scattering of electrostatic Langmuir waves
excited by two-stream instabilities, as~expected in the jet-inter-stellar-medium interaction scenario of Schlickeiser
et al. (2002)~\cite{Schlickeiser02}. 

Motivated by early $\gamma$-ray observations by the SAS-2 and COS-B satellites, already in 1979--1980, RS had considered 
inverse-Compton scattering as the dominant mechanism to produce high-energy $\gamma$-rays in astrophysical sources, pointing
out the importance of Klein-Nishina effects in the calculation of $\gamma$-ray spectra~\cite{Schlickeiser79,Schlickeiser80a,Schlickeiser80b}.
Also in leptonic models for blazars, inverse-Compton scattering by relativistic electrons in the jet is considered the dominant
high-energy emission mechanism. Target photons for Compton scattering can be the co-spatially produced synchrotron
(or electrostatic bremsstrahlung) radiation, in~which case it is termed synchrotron self-Compton (SSC) emission 
(e.g.,~\cite{Maraschi92,BM96}). 
The~first suggestion of target photon fields from outside the jet involved RS in 
two seminal papers suggesting the photon field of the accretion disk as the dominant target photon field~\mbox{\cite{DSM92,DS93}}.
Alternative sources of external target photons may be the broad-line region (BLR) (e.g.,~\cite{Sikora94}), a~dusty, infra-red
emitting torus (e.g.,~\cite{Blazejowski2000}), or~other regions of the jet (e.g.,~\mbox{\cite{GK03,TG08}}). 
The~relativistic motion
of the high-energy emission region in a blazar jet through these generally anisotropic external radiation fields leads to 
complicated transformation properties from the 
active galactic nucleus (AGN) 
rest frame into the emission-region frame, which were studied in
detail by Dermer and 
Schlickeiser in 2002~\cite{DS02}. Which of these potential radiation fields might dominate, depends 
critically on the location of the emission region, which can be constrained by the absence of obvious signatures of 
$\gamma\gamma$ absorption of high-energy and very-high-energy $\gamma$-rays by the nuclear radiation fields 
of the central AGN, with~one of the first detailed discussions of such constraints published by Dermer and
Schlickeiser in 1994~\cite{DS94}. 

The generation of the non-thermal broadband emission from blazars requires the efficient acceleration of electrons to
ultra-relativistic energies. One of the plausible mechanisms of particle acceleration acting in the relativistic jets of blazars
is diffusive shock acceleration (DSA), which was studied in the context of a general derivation of the kinetic equation of test 
particles in turbulent plasmas by RS in two seminal papers in 1989~\cite{Schlickeiser89a,Schlickeiser89b} for non-relativistic 
shock speeds, while particle acceleration by magnetic turbulence, specifically in relativistic jets was studied by Schlickeiser
and Dermer in 2000~\cite{SD00}. Particle acceleration at relativistic shocks has been considered by several authors, using both analytical
methods (e.g.,~\cite{Peacock81,KH89,Kirk00}) and Monte-Carlo techniques (e.g.,~\cite{Ellison90,BO98,ED04,NO04,SB12}).
The simulations by Niemiec and Ostrowski~\cite{NO04} and Summerlin and
Baring~\cite{SB12} indicate that diffusive shock 
acceleration at oblique, mildly relativistic shocks is able to produce relativistic, non-thermal particle spectra with a wide 
range of spectral indices, including as hard as $n(p) \propto p^{-1}$, where $p$ is the particle's~momentum. 

In two recent papers~\cite{Baring17,BB19}, we had coupled Monte-Carlo simulations of diffusive shock
acceleration (DSA), using the code of Summerlin and 
Baring~\cite{SB12}, with~time-dependent radiation transfer,
based on radiation modules originally developed by B\"ottcher, Mause and
Schlickeiser in 1997~\citep{BMS97} and
further developed as detailed in~\cite{BC02,Boettcher13}. In~those studies,  
we found that the particles' mean free path for
pitch-angle scattering, $\lambda_{\rm pas}$, which mediates the first-order Fermi process in DSA, must have a strong 
dependence on particle momentum, with~an index $\alpha > 1$ for a parameterization of $\lambda_{\rm pas} (p) \propto 
p^{\alpha}$. This likely indicates a decaying level of magneto-hydrodynamic turbulence with increasing distance from the 
shock front. Higher-energy particles, with~their larger gyro radii, then probe more distant regions from the shock
front, experiencing less efficient pitch-angle scattering. Time-dependent simulations of DSA plus 
radiation transfer
were used to fit the multi-wavelength variability of the blazars 
3C279 and Mrk 501 in~\cite{BB19} and the X-ray variability of
1ES 1959 + 650 in~\cite{Chandra21}. Multi-wavelength flares with approximately equal flare amplitude in the low-frequency 
(synchrotron) and high-frequency (Compton) components of the SED were naturally produced by an increase of the power 
injected into shock-accelerated particles, without~the need for significant changes of the plasma parameters determining 
$\lambda_{\rm pas} (p)$. 

However, an~orphan $\gamma$-ray flare on December 20, 2013, with~no significant counterpart in the synchrotron 
emission component, reported as Flare B in~\cite{Hayashida15}, presented a severe challenge to this as well as any other 
single-zone emission model for blazars. A~fit to the observed $\gamma$-ray flare was possible with a significant hardening
of the DSA-generated particle spectrum as the result of a reduction of the pitch-angle-scattering mean-free path, both in overall 
normalization $\lambda_{\rm pas} (0)$ and index $\alpha$. However, keeping the optical (synchrotron) flux approximately 
constant, as~observed, required a reduction of the magnetic field by a factor of 8.7, followed by a gradual recovery to the
quiescent-state value with a fine-tuned time dependence. While 
the authors argue that such magnetic-field reductions and subsequent gradual recoveries after the passage of a shock
have indeed been observed in interplanetary shocks (e.g.,~\cite{Baring97}), it is worth exploring alternative ways
to explain orphan $\gamma$-ray flares in blazars within the framework of the shock-in-jet model developed in
~\cite{Baring17,BB19}. 

One plausible way of producing orphan $\gamma$-ray flares in the framework of a leptonic single-zone blazar
model is the temporary enhancement of an external radiation field that serves as target for inverse-Compton
scattering. This is the basis of a class of models termed {\it synchrotron mirror models}, where the synchrotron
radiation of the high-energy emission region traveling along the jet, is reflected by a cloud to re-enter the
emission region at a later time. Such models were first considered by Ghisellini and
Madau~\cite{GM96}, however
without proper consideration of light-travel time effects, and~by B\"ottcher and 
Dermer~\cite{BD98} and Bednarek
~\cite{Bednarek98}, properly treating light-travel time effects, but~considering primarily the time-dependence of
the target-photon energy density without detailed calculations of the emerging $\gamma$-ray spectra. The~synchrotron
mirror model was more recently re-visited by Vittorini~et~al.~\cite{Vittorini14}, with~a fully time-dependent leptonic
synchrotron mirror model applied to the spectral variability of 3C454.3 in 2010 November, and~Tavani~et~al.~\cite{Tavani15}, 
considering also moving mirrors and applying the model to the light curve of the same flare B of 3C279 considered by
~\cite{BB19}. Note a similar model termed the ``ring of fire'' model by MacDonald~et~al.~\cite{MacDonald15,MacDonald17},
where the emission region passes a static synchrotron-emitting region of an outer sheath of the jet (the ``ring of fire''),
which produces very similar variability features as the synchrotron mirror~model. 

In the present paper, the~time-dependent shock-in-jet model of B\"ottcher and 
Baring~\citep{BB19} is extended to include
self-consistently a synchrotron-mirror component. Section 
~\ref{model} describes the additions to the model. Section~\ref{application}
presents the resulting spectral variability features from an attempt to apply this model to the orphan $\gamma$-ray flare
B of 3C279. Section~\ref{summary} summarizes and discusses the~results.

\section{\label{model}Model~Description}

The model developed here is a further development of the time-dependent shock-in-jet model of B\"ottcher and 
Baring~\cite{BB19}.
In addition to the radiation components already included in~\cite{BB19} , we now introduce synchrotron emission reflected by a
spherical cloud of radius $R_{\rm cl}$ at a distance $z_{\rm cl}$ from the central engine, assumed for simplicity to be located
close to the path of the jet, however, not hydrodynamically interacting with it, as~considered, e.g.,~by the jet-star interaction model
~\cite{Barkov12,Araudo13} or the cloud ablation model~\cite{Zacharias17,Zacharias19}. A~mildly relativistic, oblique shock is propagating 
along the jet, thus accelerating particles in the local environment of the shock which constitutes our moving emission region of radius $R_b$. 
The emission region is starting out at time $t_e = 0$ (in the AGN rest frame) at a height $z_0$ above the 
black-hole --- accretion-disk system 
powering the jet, and~is propagating with a bulk Lorentz factor $\Gamma$, corresponding to a speed of $\beta_{\Gamma} c$. 
Thus, at~any 
given time $t_e$, the~emission region is located at $z_e = z_0 + \beta_{\Gamma} \, c \, t_e$. 

Synchrotron radiation emitted by the emission region at $z_e$ is
reflected back by the cloud to re-enter the emission region
at a distance $z_r$ from the central engine, given by
\begin{equation} 
z_r = {2 \beta_{\Gamma} z_{\rm cl} + z_e (1 - \beta_{\Gamma}) \over 1 + \beta_{\Gamma}},
\label{zr}
\end{equation}
at a time (in the AGN rest frame) $t_r$ given by
\begin{equation}
t_r = t_e + 2 \, {z_{\rm cl} - z_e \over (1 + \beta_{\Gamma}) c}. 
\label{tr}
\end{equation}

Equation~(\ref{tr}) may be inverted to find the time at which reflected synchrotron radiation received at time $t_r$ has been emitted:
\begin{equation}
t_e = {1 + \beta_{\Gamma} \over 1 - \beta_{\Gamma}} \, t_r - {2 \over 1 - \beta_{\Gamma}} \, {z_{\rm cl} - z_0 \over c}
\label{te}
\end{equation}
implying that reflected synchrotron emission will be received starting at a time $t_0$ (corresponding to $t_e = 0$) given by
\begin{equation}
t_0 = {2 \over 1 + \beta_{\Gamma}} \, {z_{\rm cl} - z_0 \over c}. 
\label{t0}
\end{equation}

Reflected synchrotron radiation will be received by the emission region until it passes the cloud at $t_{\rm pass} \approx (z_{\rm cl} - z_0)/
(\beta_{\Gamma} c)$. 

The code writes out the observed synchrotron emission spectra, 
$\nu F_{\nu}^{\rm sy} (t_e)$ for every time step (with times in the 
observer's frame) as the shock propagates along the jet. Therefore, at~any time $t_{\rm AGN} > t_0$, 
one can use Equation~(\ref{te}) to 
find the time (in the AGN frame) at which synchrotron radiation reflected back into the emission region, has been emitted. The~
synchrotorn flux irradiating the cloud, $\nu F_{\nu}^{\rm cl}$, is then found as
\begin{equation}
\nu F_{\nu}^{\rm cl} = \nu F_{\nu}^{\rm sy} (t_e) \, {d_L^2 \over (z_{\rm cl} - z_e)^2}
\label{Fcl}
\end{equation}
where $d_L$ is the luminosity distance to the source. 

Assuming, for~simplicity, that the cloud re-radiates a fraction $\tau_{\rm cl}$ of the impinging synchroton radiation isotropically, it will 
emit a spectral luminosity of
 $\nu L_{\rm nu}^{\rm cl} = \pi R_{\rm cl}^2 \, \tau_{\rm cl} \, \nu F_{\nu}^{\rm cl}$. The~emission 
region will thus receive a flux of Reflected Synchrotron 
(RS --- happy coincidence) radiation, in~the comoving frame, of~\begin{equation}
\nu' F_{\nu'}^{\rm RS} (t_r) \approx {R_{\rm cl}^2 \, \tau_{\rm cl} \, \nu F_{\nu}^{\rm sy} (t_e) \Gamma^4 \, d_L^2 \over 4 \,
(z_{\rm cl} - z_e)^2 \, (z_{\rm cl} - z_r)^2}
\label{FRS}
\end{equation}
where $\nu' \approx \Gamma \nu$ is the photon frequency in the co-moving~frame. 

The code evaluates a time-evolving reflected-synchrotron photon field in the emission region, $n'_{\rm RS} (\epsilon', t_r')$, where 
$\epsilon' = h \nu'/(m_e c^2)$ is the dimensionless photon energy in the emission-region frame by~the 
interplay of RS emission entering the emission region at a rate $dn'_{\rm RS, inj} (\epsilon', t'_r)/dt'_r = \pi \, R_b^2 \nu' F^{\rm RS}_{\nu'}
(t_r)/(V_b \, {\epsilon'}^2 \, m_e c^2)$, where $V_b$ is the volume of the emission region, and~escape on an escape time scale $t'_{\rm esc} 
= 3 \, R_b/(4 \, c)$, over~a simulation time step $\Delta t'$ as
\begin{equation}
\Delta n'_{\rm RS} (\epsilon', t'_r) = \left( {\pi R_b^2 \over V_b} \, {\nu' F_{\nu'}^{\rm RS} (t_r) \over {\epsilon'}^2 \, m_e c^2}  - {n'_{\rm RS}
(\epsilon', t'_r) \over t'_{\rm esc}} \right) \, \Delta t'.
\label{dnRS}
\end{equation} 
In the above expressions, $h$ is the Planck constant, $m_e$ the electron mass, and~$t'_r = t_r/\Gamma$.

The time-dependent RS photon field resulting from Eq. \ref{dnRS} acts target for inverse-Compton scattering to produce the synchrotron-mirror 
Compton emission, and~synchrotron-mirror Compton cooling is included self-consistently. For~the evaluation of the synchrotron-mirror Compton 
emission, it is assumed, for simplicity, that the target photons enter the jet directly from the front, and~the head-on approximation 
(e.g.,~\cite{DM09}) for the Compton 
cross section is used. Hence, photons scattered along the viewing direction, making an angle $\theta'_{\rm obs}$ with respect to the jet axis
in the co-moving frame of the emission region, with~$\mu'_{\rm obs} \equiv \cos\theta'_{\rm obs}$, have been scattered by a scattering angle $\mu'
= - \mu'_{\rm obs}$. The~rate density at which Reflected  Synchrotron Compton (RSC) emission is produced, is calculated as

\begin{equation}
{\dot{n}'}_{\rm RSC} (\epsilon'_s, \Omega'_s) = c \, \int\limits_0^{\infty} d\epsilon' \, n'_{\rm RS} (\epsilon') \int\limits_1^{\infty} d\gamma \,
(1 + \beta \, \mu'_{\rm obs}) \, n_e (\gamma) \, {d\sigma_C \over d\epsilon'_s} (\epsilon', \epsilon'_s, \mu')
\label{ndotCRS}
\end{equation}
where
\begin{equation}
{d\sigma_C \over d\epsilon'_s} = {\pi \, r_e^2 \over \gamma \, \epsilon_e} \left\lbrace y + {1 \over y} - {2 \, \epsilon'_s \over \gamma \, \epsilon_e \, y}
+ \left( {\epsilon'_s \over \gamma \, \epsilon_e \, y} \right)^2 \right\rbrace
\label{dsigma}
\end{equation}
with $\epsilon_e = \epsilon' \, \gamma \, (1 + \beta \, \mu'_{\rm obs})$, $y = 1 - \epsilon'_s/\gamma$ and $\beta = \sqrt{1 - 1/\gamma^2}$~\cite{DM09}.

\section{\label{application}Results: Spectral Variability~Features}

As detailed in the introduction, the~study of the synchrotron mirror model developed here was motivated by the difficulties
in modeling the orphan $\gamma$-ray flare B of 3C279 in December 2013 reported by Hayashida~et~al.~\cite{Hayashida15}. 
We therefore start with the quiescent-state parameters of the shock-in-jet model for 3C279 used in~\cite{BB19}. The~emission
region is set to start out at $z_0 = 0.1$~pc, and~the cloud acting as the mirror is assumed to be located at $z_{\rm cl} =
1$~pc. The~bulk Lorentz factor is $\Gamma = 15$, as~used in~\cite{BB19}. The~cloud radius is assumed to be $R_{\rm cl}
= 3 \times 10^{17}$~cm and its reflective fraction is $\tau_{\rm cl} = 0.1$. The~complete list of model parameters can
be found in Table~\ref{model_parameters}.

\begin{specialtable}[H]
\caption{\label{model_parameters}Relevant model parameters for the case study motivated by the December 2013 flare of~3C279.}
\begin{tabular*}{\hsize}{@{}@{\extracolsep{\fill}}lcc@{}}

\toprule
\textbf{Parameter} & \textbf{Symbol} & \textbf{Value} \cr
\midrule
Electron injection luminosity & $L_{\rm inj}$ & $1.0 \times 10^{43}$~erg~s$^{-1}$ \\ 
Pitch-angle mean free-path (m.f.p.) scaling normalization & $\eta_0$ & 100 \\  
Pitch-angle m.f.p. scaling index & $\alpha$ & 3.0 \\
Magnetic field & $B$ & 0.8~G \\
Electron escape time scale factor & $\eta_{\rm esc}$ & 3.0 \\
Emission region radius & $R_b$ & $2.0 \times 10^{16}$~cm \\
Bulk Lorentz factor & $\Gamma$ & 15 \\
Viewing angle & $\theta_{\rm obs}$ & 3.82$^o$ \\
Initial distance of the shock from the black hole (BH) along the jet & $z_0$ & 0.1~pc \\ 
Distance of the cloud from the BH & $z_{\rm cl}$ & 1~pc \\
Radius of the cloud & $R_{\rm cl} $ & $3 \times 10^{17}$~cm \\
Reflective fraction of the cloud & $\tau_{\rm cl}$ & 0.1 \\  
Mass of the BH & $M_{\rm BH}$ & $5 \times 10^8 \, M_{\odot}$ \\
Luminosity of the accretion disk & $L_d$ & $6 \times 10^{45}$~erg~s$^{-1}$ \\
Black-body temperature of external radiation field & $T_{\rm ext}$ & 300~K \\
Energy density of external radiation field & $u_{\rm ext}$ & $4 \times 10^{-4}$~erg~cm$^{-3}$ \\
\bottomrule
\end{tabular*}

\end{specialtable}

\begin{figure}[H]
\includegraphics[width=12cm]{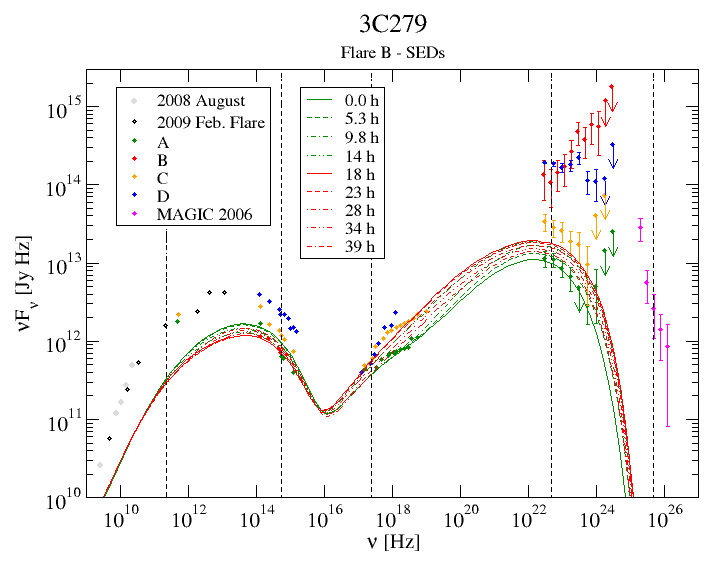}
\caption{\label{FlareBSEDs} Spectral energy distributions (SEDs)  of 3C279 in 2013--2014, from~\cite{Hayashida15}, along 
 with snap-shot model SEDs from the shock-in-jet
synchrotron-mirror model. The~dashed vertical lines indicate the frequencies at which light curves and  hardness-intensity relations 
were extracted. The legend follows the nomenclature of different periods from Hayashida et al. (2015) \cite{Hayashida15}. 
 }
\end{figure}

The resulting sequence of snap-shot SEDs
(starting right before the onset of the synchrotron-mirror Compton emission) is illustrated in Figure~\ref{FlareBSEDs}. 
It is clear that the model does produce a significant orphan $\gamma$-ray flare, accompanied by a slight reduction 
of the synchrotron emission due to the increased Compton cooling of relativistic electrons. The~latter is consistent with
the observed evolution of the SED. However, the~amplitude of the orphan $\gamma$-ray flare amounts only to an 
enhancement of the $\gamma$-ray flux by a factor of $\sim$2, in~contrast to the observed dramatic increase by
a factor of $\gtrsim$10. Even increasing the synchrotron-mirror efficiency (e.g., by~increasing $\tau_{\rm cl}$ or
$R_{\rm cl}$; see Equation~(\ref{FRS})), does not increase the $\gamma$-ray flare amplitude substantially. The~reason 
for this is that the $\gamma$-ray flux is limited by the available power injected into shock-accelerated electrons,
which are already in the fast-cooling regime, thus radiating very efficiently. Any further enhancement of the 
reflected-synchrotron energy density will only suppress the synchrotron emission further, but~not lead to a
significant increase of the $\gamma$-ray flare amplitude. We therefore conclude that a pure shock-in-jet 
synchrotron mirror scenario is not able to produce the observed large-amplitude orphan $\gamma$-ray flare
in 3C279 in December 2013. In~order to achieve this, additional power would need to be injected into shock-accelerated
electrons, leaving us with the same difficulties encountered in~\cite{BB19}, i.e.,~requiring a fine-tuned reduction and
gradual recovery of the magnetic~field.

\begin{figure}[H]
\includegraphics[width=12cm]{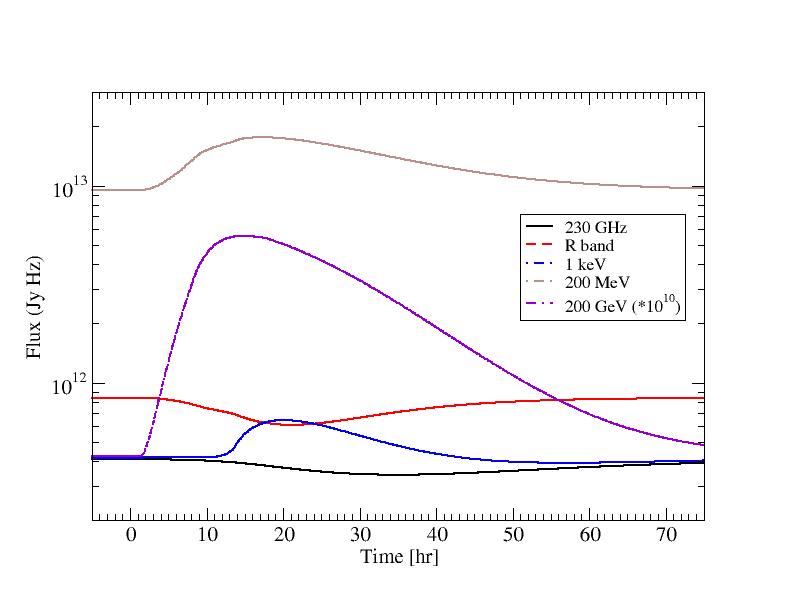}
\caption{\label{FlareBlcs}Model light curves in various frequency/energy bands resulting from the synchrotron mirror
simulation illustrated in Figure~\ref{FlareBSEDs} at the 5 representative frequencies/energies marked by the vertical dashed lines. 
Note that the very-high-energy (VHE, 200 GeV)  $\gamma$-ray flux is scaled up by a factor of $10^{10}$ in order to be visible on the plot. 
}
\end{figure}

Nevertheless, in~spite of its inapplicability to this particular orphan flare, it is worthwhile considering this simulation for a
generic study of the expected spectral variability patterns in the shock-in-jet synchrotron mirror model. The~multi-wavelength
light curves at 5 representative frequencies (high-frequency radio, optical, X-rays, high-energy [HE, 200~MeV], and 
very-high-energy [VHE, 200 GeV] $\gamma$-rays) are shown in Figure~\ref{FlareBlcs}.
All light curves in the Compton SED component (X-rays to VHE $\gamma$-rays) show a flare due to the synchrotron-mirror Compton
emission. Note that the VHE $\gamma$-ray light curve had to be scaled up by a factor of $10^{10}$ to be visible on this plot. Thus,
the apparently large VHE flare is actually at undetectably low flux levels for the parameters chosen here. In~contrast, the~230~GHz 
radio and optical light curves show a dip due to increased radiative cooling during the synchrotron mirror action. The~radio dip is 
significantly delayed compared to the optical due to the longer cooling time scales of electrons emitting in the radio~band. 

\vspace{-9pt}

\begin{figure}[H]
\includegraphics[width=12cm]{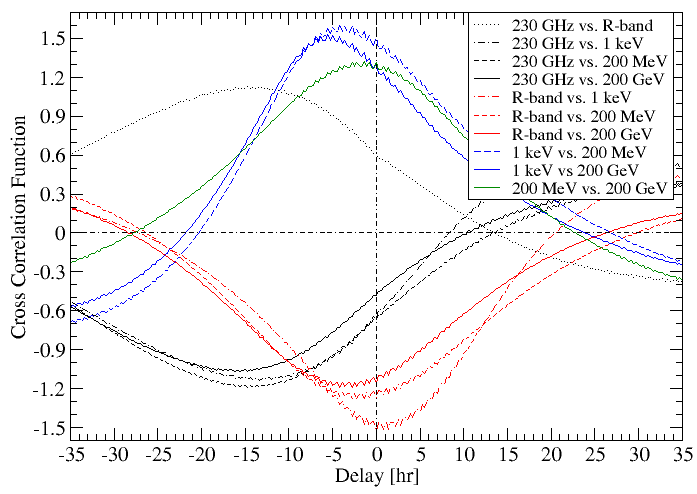}
\caption{Cross-correlation functions between the model light curves in various energy/frequency~bands.}
\label{FlareBDCFs}
\end{figure}

Cross-correlation functions between the various light curves from Figure~\ref{FlareBlcs} are shown in Figure~\ref{FlareBDCFs}. As~expected from inspection of the light curves, significant positive correlations between X-rays and the 2 $\gamma$-ray bands
with only small time lags ($\gamma$-rays leading X-rays by a few hours) and between the radio and optical band, with~optical
leading the radio by $\sim$15~h, are seen. The~synchrotron (radio and optical) light curves are anti-correlated with the
Compton (X-rays and $\gamma$-rays) ones, again with a significant lag of the radio emission by $\sim$15~h. 

\begin{figure}[H]
\includegraphics[width=12cm]{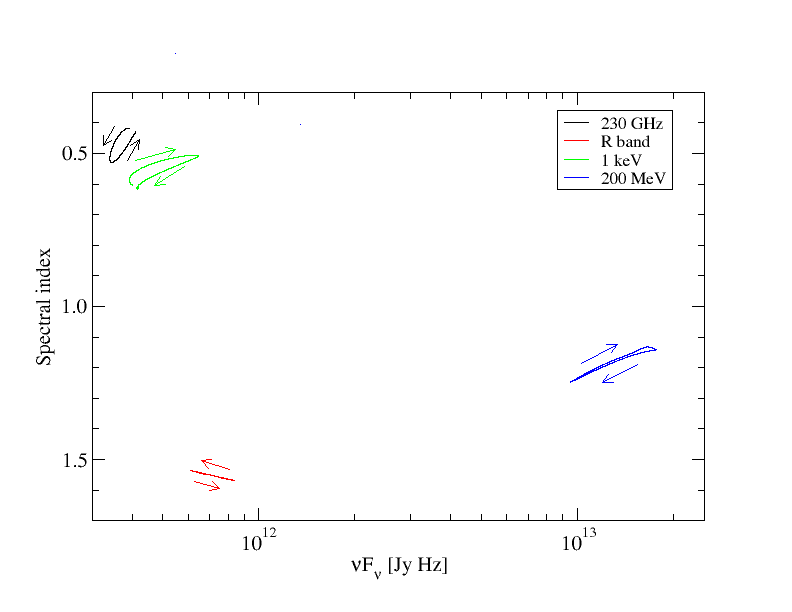}
\caption{\label{FlareBHIDs}Model hardness-intensity diagrams at the selected frequencies/energies. The~spectral index $a$ is
defined by $F_{\nu} \propto \nu^{-a}$, so that a smaller value indicates a harder spectrum. The~VHE band has been omitted 
here due to its unobservably low flux level and very steep local spectral index. Arrow indicate the evolution in time.   
}
\end{figure}

Figure~\ref{FlareBHIDs} shows the hardness-intensity diagrams for the 5 selected frequencies/energies, i.e.,~the evolution of
the local spectral index ($a$, defined by $F_{\nu} \propto \nu^{-a}$) vs. differential flux. Generally, all bands, except~the optical, 
exhibit the frequently observed harder-when-brighter trend. Only the radio and X-ray bands show very moderate spectral hysteresis. 
The dip in the optical R-band)
 light curve is accompanied by a very slight redder-when-brighter trend, likely as the consequence of 
the increased radiative cooling during the synchrotron-mirror~action. 

\section{\label{summary}Summary, Discussion, and~Conclusions} 

In this paper, the leptonic shock-in-jet blazar model of~\cite{BB19} is extended with the addition of a self-consistent synchrotron mirror
component. This was motivated by the difficulty in modeling orphan $\gamma$-ray flares with such an effectively single-zone model.
A particularly high-amplitude (factor of $\sim$10) orphan $\gamma$-ray flare of the blazar 3C279 from December 2013 was chosen as a 
case 
study. However, the~attempt to model this flare with the shock-in-jet synchrotron mirror model developed here, failed because the
maximum $\gamma$-ray flux was limited by the (fixed) amount of power injected into shock-accelerated electrons, allowing for 
orphan flares with amplitudes of at most $\sim$2--3. Higher-amplitude flares would require an enhanced energy injection into
relativistic electrons, in~addition to more efficient pitch-angle scattering, leading to a harder electron spectrum. However, this would
cause the same difficulties of having to decrease the magnetic field, followed by a fine-tuned recovery to its quiescent state, as~were
encountered in~\cite{BB19}. 

More successful model representations of this particular flare of 3C279 were presented by several authors. Hayashida~et~al.~\cite{Hayashida15}
use the model of Nalewajko~et~al.~\cite{Nalewajko14} to reproduce this orphan $\gamma$-ray flare by introducing an extreme hardening
of the electron spectrum, along with a location of the emission region much closer to the BH and accretion disk. A~hard electron spectrum
$n_e (\gamma) \propto \gamma^{-1}$ up to a cut-off energy of a few 1000 is invoked, which may be difficult, but~not impossible, to~achieve 
with standard particle acceleration mechanisms. Asano and
Hayashida~\cite{AH15} employ a time-dependent one-zone model with second-order
Fermi acceleration, where an enhanced acceleration efficiency leads to a hardening of the electron spectrum, and~a significant reduction
of the magnetic field is required to suppress a simultaneous optical flare. While their model represents the $\gamma$-ray spectrum during 
the flare well, it does predict a non-negligible optical synchrotron flare accompanying the $\gamma$-ray flare. A~similar strategy, based on
an analytical solution to the steady-state electron distribution, was adopted by Lewis~et~al.~\cite{Lewis19}, also requiring a significant reduction
of the magnetic field to suppress a simultaneous optical synchrotron flare. Yan~et~al.~\cite{Yan16} modelled the orphan-flare SED using a 
time-dependent single-zone model with rapid electron cooling. However, it is unclear whether a transition from the quiescent to this flaring 
state may be produced in a natural way. Lepto-hadronic models naturally
de-couple the (proton-initiated) high-energy emission from the (electron-initiated) synchrotron radiation and therefore offer an alternative 
way of reproducing orphan $\gamma$-ray flares. Paliya~et~al.~\cite{Paliya16} used the time-dependent lepto-hadronic model of Diltz
et~al.~\cite{Diltz15} to model the December 2013 orphan $\gamma$-ray flare of 3C279. They also considered the possibility of a two-zone
model, with~a small emission region emitting an SED with very large Compton dominance, emerging during the time of the orphan $\gamma$-ray
flare. 

A representative simulation of the shock-in-jet synchrotron mirror scenario was then used for a generic study of the expected spectral
variability patterns. X-ray and $\gamma$-ray light curves as well as radio and optical light curves are significantly correlated with each
other, while radio and optical light curves are significantly anti-correlated, with~radio and optical (synchrotron) dips accompanying the
high-energy flare resulting from more efficient radiative cooling of the electrons. The~response in the radio light curves is found to be
delayed by $\sim$15~h with respect to other bands. In~the scenario investigated here, where no changes to the diffusive shock
acceleration process along the entire evolution of the flare are assumed, significant spectral hysteresis is not expected, but~a mild 
harder-when-brighter trend in most wavebands is~found.  

While it is found that the specific December 2013 orphan $\gamma$-ray flare of 3C279 can not be successfully reproduced with
this scenario, it may be applicable to other, more moderate orphan $\gamma$-ray flares. Especially the expected anti-correlation
between Compton and synchrotron wavebands may serve as a smoking-gun signature of this scenario.  The~failure of the shock-in-jet 
synchrotron mirror model for the December 2013 flare of 3C279 was primarily caused by the fact that the shock-accelerated, relativistic
electrons were already in the fast-cooling regime and radiating very efficiently, as~was required by the fit to the quiescent state of 3C279.
If the quiescent emission of a blazar can be produced by a less radiatively efficient configuration, then the increase in radiative efficiency
in the synchrotron mirror scenario may lead to substantially higher-amplitude flares. A~systematic study of different scenarios and 
applications to other sources will be presented in a forthcoming~publication. 

\vspace{6pt} 

\funding{The work of M.B. is supported by the South African Researc Chairs Initiative of the National Research Foundation\footnote{Any opinion, 
finding, and~conclusion or recommendation expressed in this material is that of the authors, and~the NRF does not accept any liability in this~regard.}
and the Department of Science and Innovation of South Africa through SARChI grant no. 64789.  
}

\conflictsofinterest{The author declares no conflict of interest.}
\end{paracol}

\reftitle{References}




\end{document}